\documentstyle[12pt,epsf]{article}
\begin{document}

\begin{center}
{\bf VOLUME FREE ELECTRON LASERS}

\vspace{1pt}

{\bf V.G.Baryshevsky}
\end{center}

{\bf \vspace{1pt}}

\begin{center}
Nuclear Problems Institute, Bobruiskaya Str.11, Minsk 220080 Belarus

Electronic address: bar@inp.minsk.by

\vspace{1pt}

{\bf 1.Introduction}
\end{center}

\vspace{1pt}

There has been recently considerable theoretical and experimental
interest in the concept of free electron lasers (FELs) \lbrack 1,2\rbrack .
It has been shown that free
electron lasers can operate due to different radiation processes: "magnetic
bremsstrahlung" in the undulator, Smith-Purcel and Cherenkov radiations,
radiation in the laser wave.
Independing on spontaneous radiation mechanism being principle for a
definite FEL scheme, all existing FEL devices use for the feedback forming
either two parallel  mirrors placed at the ends of the working area or
one-dimentional diffraction grating in which transmitted and diffracted
(reflected) waves propagate along the electron beam velocity direction
(one-dimentional distributed feedback  (DFB)) (Fig.1,2).

\begin{figure}[h]
\epsfxsize = 7.8 cm
\centerline{\epsfbox{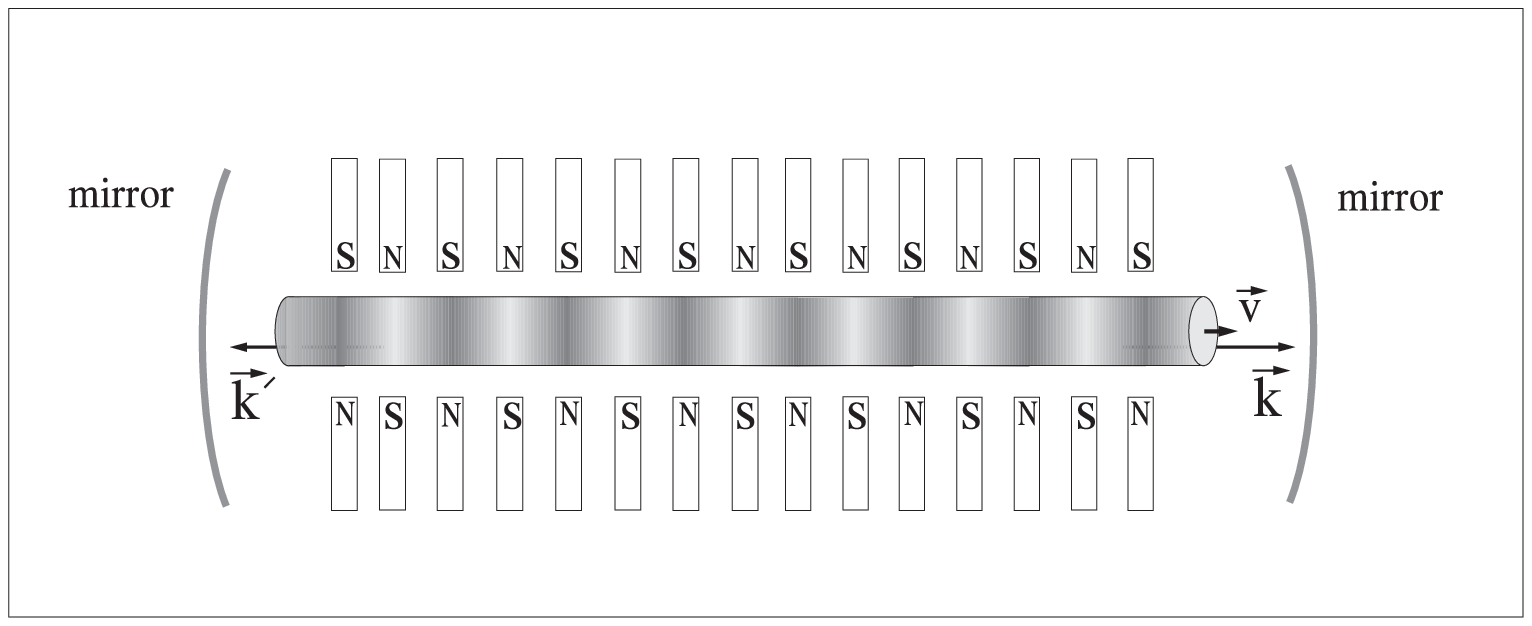}}
\caption{Free electron laser.}
\end{figure}

\begin{figure}[h]
\epsfxsize = 7.8 cm
\centerline{\epsfbox{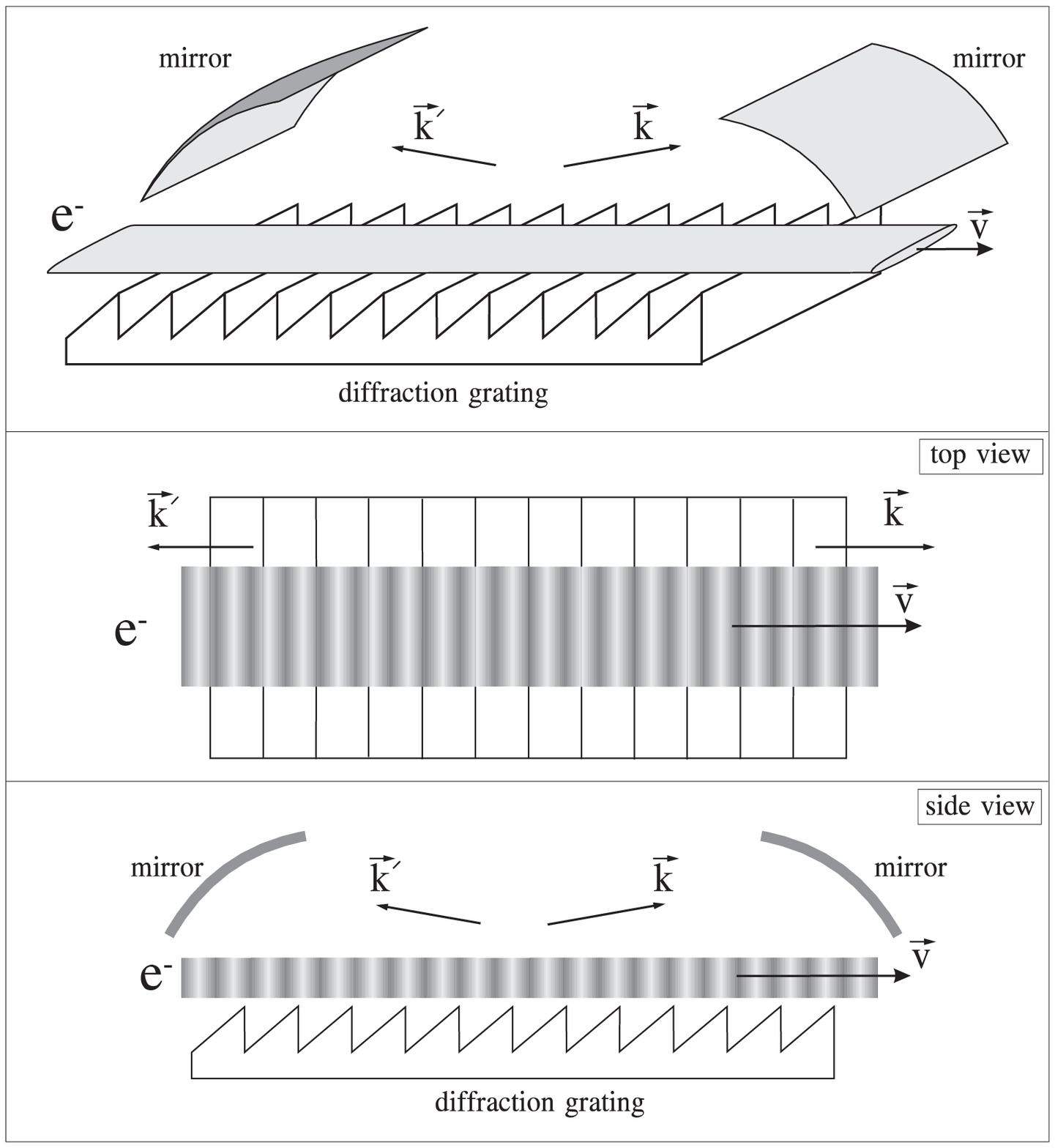}}
\caption{Smith-Purcel FEL (orotron).}
\end{figure}

According \lbrack 3\rbrack,
the dispersion equation of the FEL
in the collective interaction regime is reduced to
that of the conventional travelling wave amplifier \lbrack 4\rbrack\
and the FEL gain at the conditions of synchronism  is proportional
to $\rho _{o}^{1/3}$,where $\rho _{o}$ is the density of the electron beam.

The volume FEL  has been suggested as one of the alternative schemes
of  FEL which provides possibility to design compact sources in various
spectral ranges including ultraviolet and X-ray  \lbrack 5-10\rbrack .

The main peculiarity of VFEL is the use of one, two or three-dimensional
grating as a
volume resonator providing three-dimensional distributed feedback
(Fig.3-5).

\begin{figure}[h]
\epsfxsize = 7.8 cm
\centerline{\epsfbox{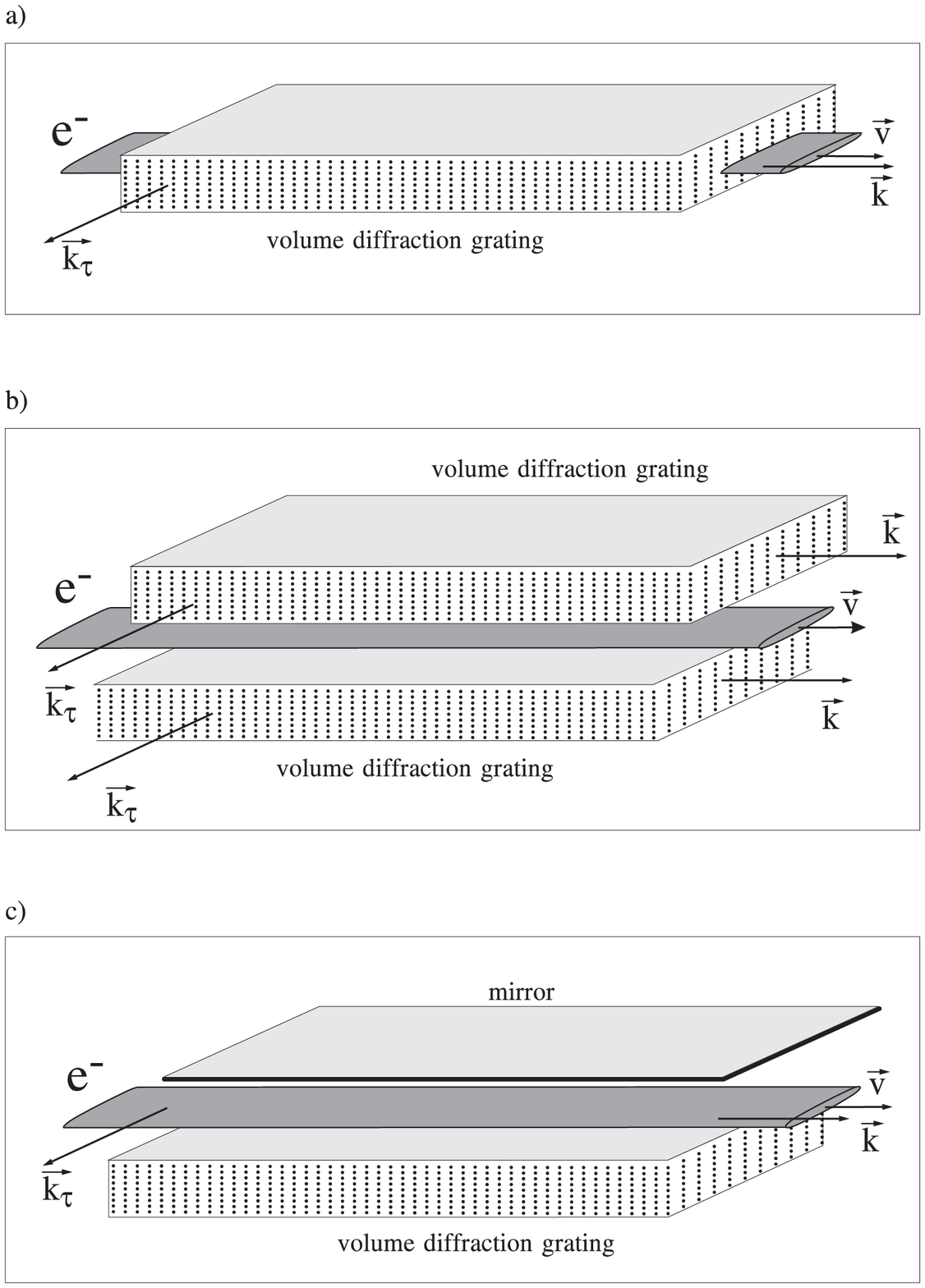}}
\caption{Parametric quasi-Cherenkov VFEL.}
\end{figure}

\begin{figure}
\epsfxsize = 7.8 cm
\centerline{\epsfbox{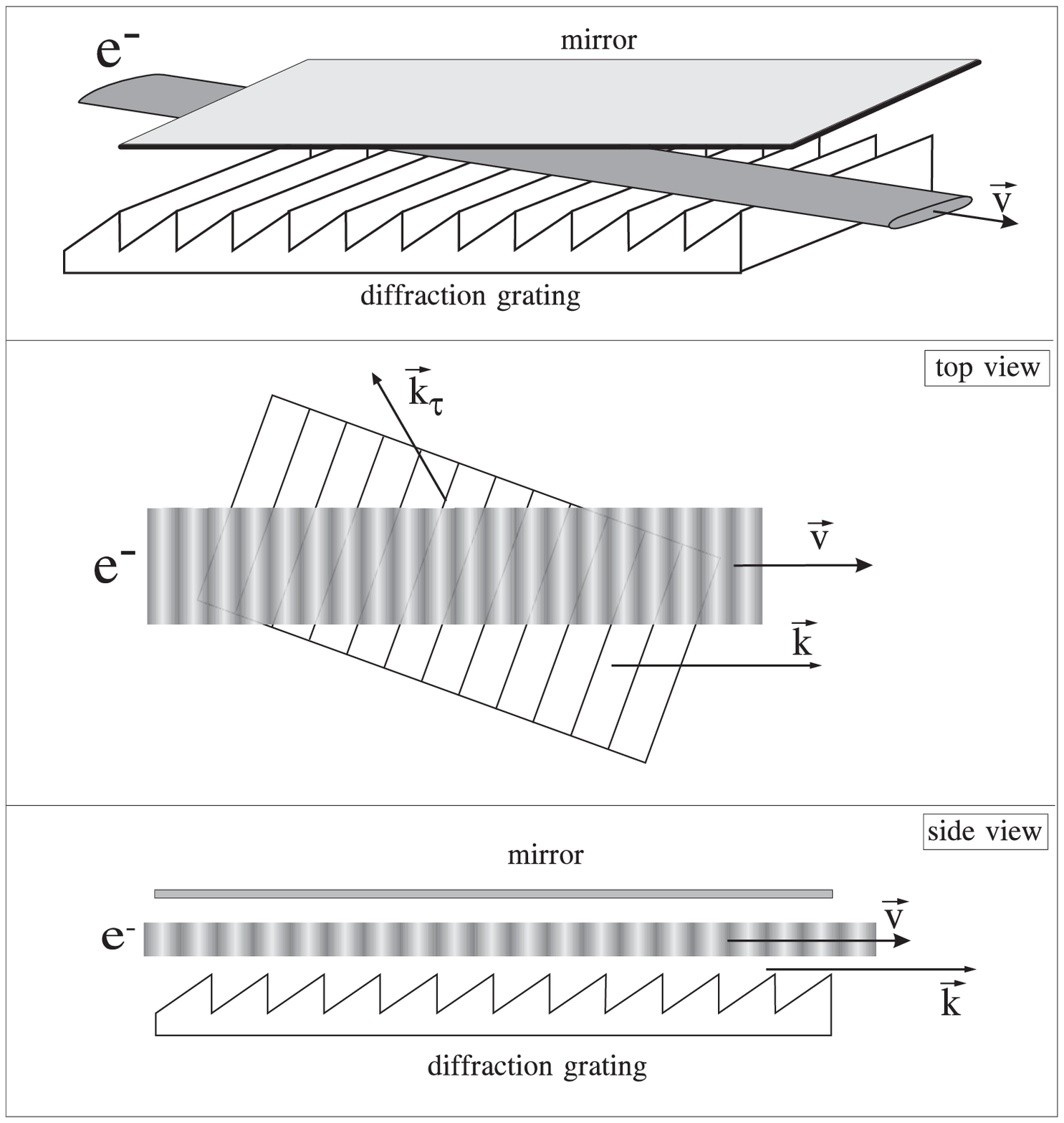}}
\caption{Parametric quasi-Cherenkov VFEL.}
\end{figure}

\begin{figure}
\epsfxsize = 7.8 cm
\centerline{\epsfbox{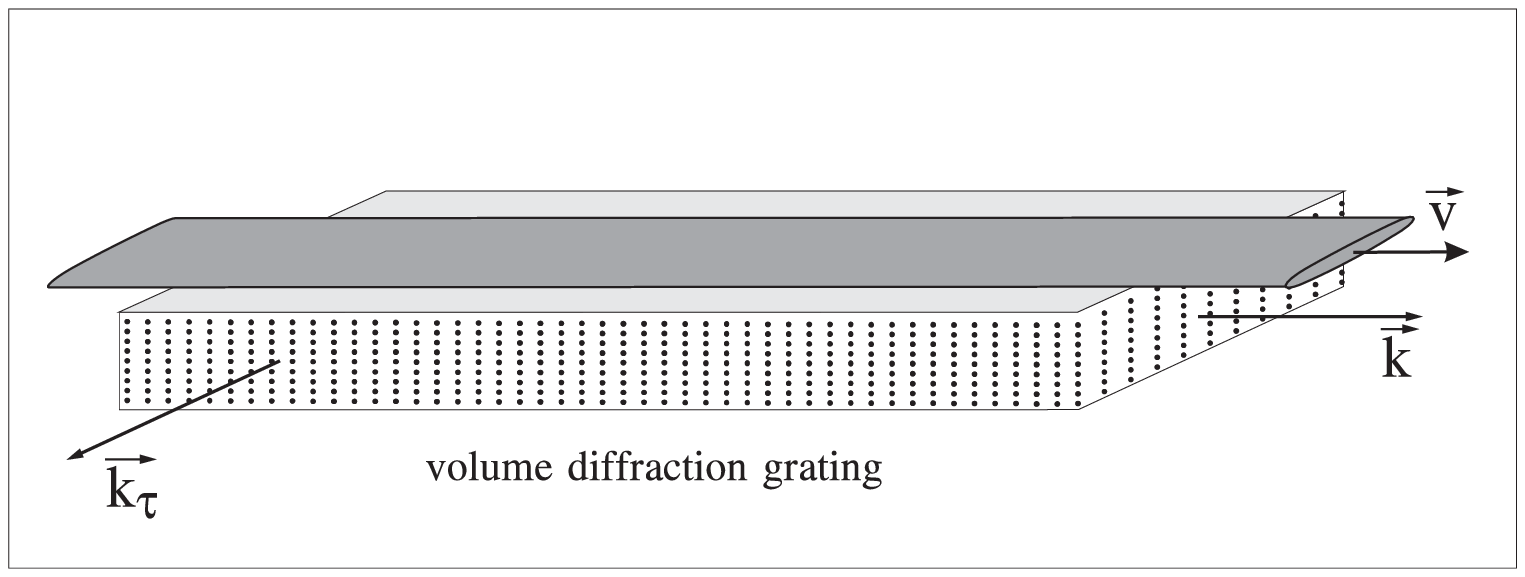}}
\caption{Surface VFEL.}
\end{figure}

It is important to emphasize that even one-dimensional diffraction
grating may provide non-one-dimensional (volume) feedback if the diffracted
wave moves in nonback direction (the Bragg diffractive angle does not equal
$\frac{\pi}{2}$).
This results in essential modification of the
VFEL gain and lasing processes
providing,
under specific conditions, more effective radiation process
as compared
with conventional FELs using one-dimensional distributed feedback.
The VFEL gain
at the conditions of synchronism  is proportional to $\rho _{0}^{1/S+1}$
where $S$
is the number of diffracted waves. Volume FEL, if realised, could be made with
much more compact device structure compared with the conventional FEL
and therefore, may be
interesting for applications in different wavelength regions: from
submillimeter to $X$-ray \lbrack 5-9\rbrack.

It should be emphasized that a fast
destruction of the synchronism condition between a
particle and an emitted electromagnetic wave is characteristic for
the VFEL scheme with an electron
beam passing through a diffraction grating.
This leads
to the essential increase of the generation threshold parameters. The
reduction of the influence of multiple scattering becomes possible when
electron beam moves either in the split of a grating (vacuum VFEL)
or over a surface of
a grating (surface FEL, SFEL) at a distance $d\leq \lambda \gamma $
(Fig.3-5)($\lambda $ is the photon wave length, $\gamma $ is the Lorentz
factor) \lbrack 10\rbrack . The SFEL has been studied in \lbrack 8\rbrack .

It is easy to understand that vacuum VFEL turns into SFEL when width of the
grating split grows.

Radiation mechanisms being the basis of VFEL and SFEL can be various
(Cherenkov, Smith-Purcel and so on).
The spontaneous surface parametric radiation (SSPR) \lbrack 10\rbrack\
may be used for SFEL, for example.
We should distinguish the SSPR from Smith-Purcel radiation \lbrack 11,12\rbrack.

The difference between these two types of radiation may be shown
by analysing the radiation frequency dependence on electron energy.
In the case of Smith-Purcel radiation the photon frequency is proportional
to:
$$
\omega \sim \frac 1{ \frac 1{\gamma^2}+\theta^2}
$$
and, for photons emitted at  small angles to the electron velocity,
the radiation frequency depends on the electron energy as $\gamma^2$.
Moreover, this wave propagates in vacuum.

\noindent
In the case of SSPR, the frequency of photons emitted even at small angles
to the electron velocity does not practically depend on the electron
energy but is determined by the Bragg condition. This radiation propagates
inside the grating and leaves it only through a grating-vacuum boundary.
The microscopic nature  of both types of radiation is similar: they
are stipulated by the medium atoms polarisation caused by an electromagnetic
field of a moving charged particle.

\begin{figure}[h]
\epsfxsize = 7.8 cm
\centerline{\epsfbox{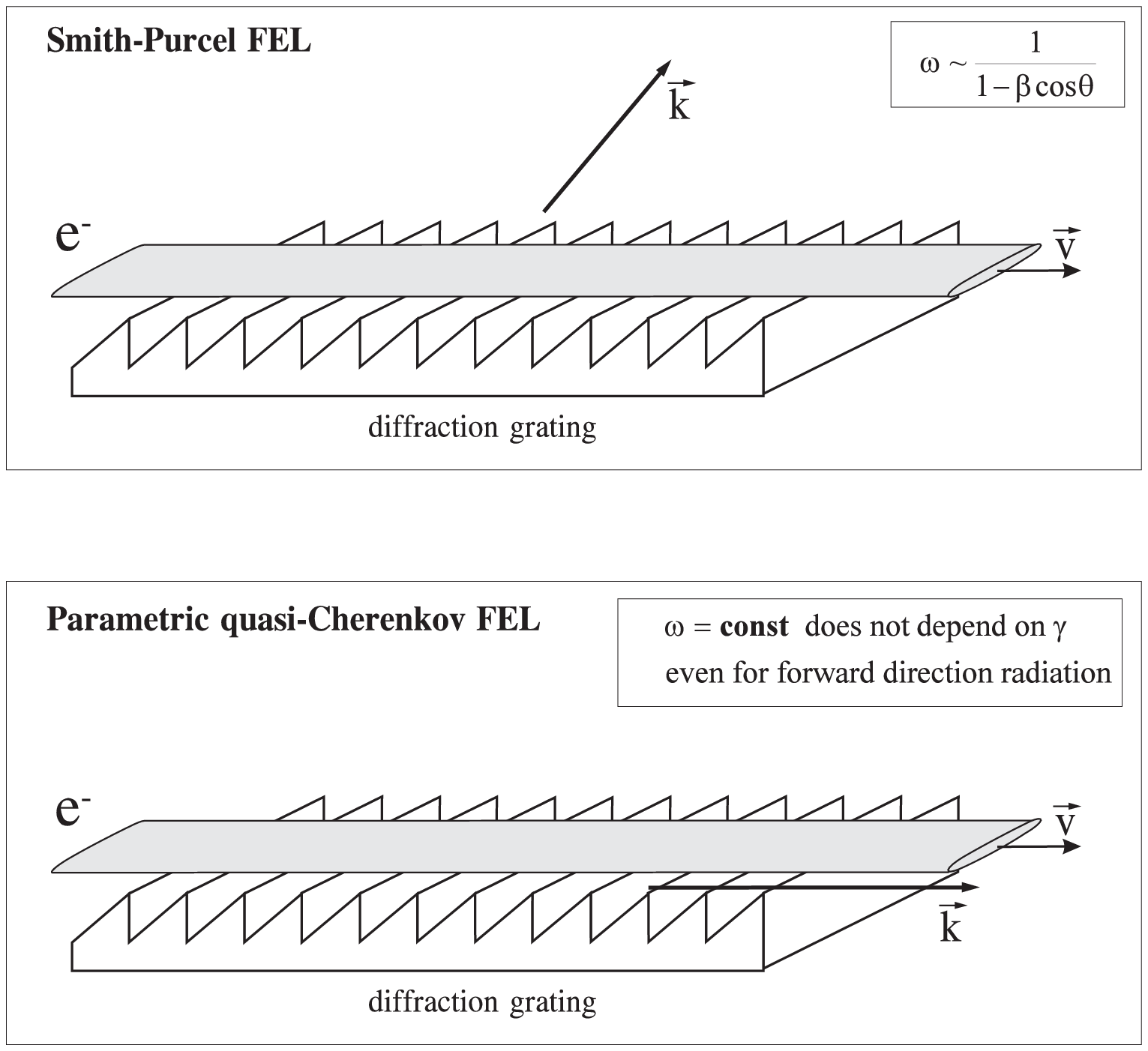}}
\caption{Smith-Purcel FEL and parametric quasi-Cherenkov VFEL comparison. }
\end{figure}

In the Smith-Purcel FEL (orotron, diffraction radiation generator)
(see \lbrack 13-17\rbrack\ ) an
electron beam passes over a reflecting diffraction grating,
two mirrors (or diffraction grating) are used for one-dimensional
feedback forming (Fig.2).

In the VFEL the non-one-dimensional
feedback forming by the diffraction grating is used (Fig.3-5).

In the present paper the equations describing the VFEL lasing in
case of an electron beam moving either in a split of a diffraction
grating or in
a vacuum waveguide containing a diffraction grating (vacuum
VFEL)(Fig.3,4) have been obtained. The dispersion equation allowing
to find the
vacuum VFEL gain in one-mode generation regime have been considered.

\begin{center}
{\bf 2. Basic formulas describing vacuum VFEL lasing.}
\end{center}

The interaction of an electron beam and an electromagnetic wave propagating
along a waveguide in the vacuum VFEL is described by the Maxwell and
electron movement equations
\begin{equation}
curl~curl~\vec E (\vec r, \omega)-\frac{\omega^2}{c^2}
\varepsilon (\vec r,\omega )\vec E (\vec r,\omega )= \frac{{4 \pi
i\omega }}{c^{2}}{\vec j (\vec{r},\omega )}
\label{eqn.1}
\end{equation}
\begin{equation}
div~\varepsilon (\vec{r},\omega )\vec{E}(\vec{r},\omega )=4\pi \rho (\vec{r}%
,\omega )
\end{equation}
\begin{equation}
-i\omega \rho (\vec{r},\omega )+div{\vec{j}(\vec{r},\omega )}=0
\label{eqn.3}
\end{equation}
where $\vec{E}(\vec{r},\omega )=\int e^{i\omega t}\vec{E}(\vec{r},t)dt$
is the Fourier transformation of the electric field $\vec{E}(\vec{r},t)$;
$\varepsilon (\vec{r},\omega )$
is the dielectric susceptibility of the
diffraction grating;
$\vec j (\vec r,\omega)$ and $\rho (\vec r,\omega )$  are the Fourier
transformations of the electric current density $\vec j (\vec r,t)$
and electric charge
density of the beam $\rho(\vec{r},t)$, respectively.
\begin{equation}
\vec{j}(\vec{r},t)=e\sum_{\alpha }{\vec v}_{\alpha }(t)\delta \left( \vec{r}-%
\vec{r}_{\alpha }(t)\right)
\label{eqn.4}
\end{equation}
\begin{equation}
\rho (\vec{r},t)=e\sum_{\alpha }\delta \left( \vec{r}-\vec{r}_{\alpha
}(t)\right)
\label{eqn.5}
\end{equation}
\vspace{1pt}
$\vec{r}_{\alpha}, \vec v_{\alpha }(t)$ are electron radius-vector and
velocity. The subscript $\alpha $ denotes the particle's number.

Movement equations can be written in the form
\begin{equation}
\frac{d {\vec v}_{\alpha }(t)}{dt}=\frac{e}{m\gamma }\left\{ \vec{E}(\vec{r}_{\alpha
}(t),t)+\frac{1}{c}\left[ \vec{v}_{\alpha }\left( t\right) \times \vec{H}((%
\vec{r}_{\alpha }(t),t)\right] -\frac{\vec{v}_{\alpha }}{c^{2}}(\vec{v}%
_{\alpha }\vec{E}(\vec{r}_{\alpha }(t),t)\right\},
\label{eqn.6}
\end{equation}
where $\vec{E}(\vec{r}_{\alpha }(t),t)~$and $\vec{H}(\vec{r}_{\alpha }(t),t)~
$are the electric field and the magnetic field of the electromagnetic wave
in the point $\vec{r}_{\alpha }(t)$ at the time moment $t$,
$\gamma=(1-\frac{v_\alpha^2}{c^2})^{-\frac12}$.

Let us consider a sheet electron
beam passed over a diffraction grating placed in a plane waveguide. At
first view this generator is similar to Smith-Purcel FEL (orotron or
diffraction radiation generator) \lbrack 3, 13-16\rbrack .
But, in the volume FEL the radiated wave wavelength $\lambda$ is of the
same
order as the diffraction grating period, the wave undergoes Bragg
diffraction on the Bragg angle non-equal to $\frac{\pi}{2}$
and the diffraction grating provides the volume distribution feedback.
Let the $(y,z)$
coordinate plane be parallel to the waveguide (diffraction grating) surface.
In the absence of the electron beam the current $\vec{j}=0$ and the
density $\rho=0$.
Equations (1, 2) become periodic in $y,z$ directions (they are not
periodic in $x$-direction). In this case the
waveguide dielectric susceptibility can be written as:
\begin{equation}
\varepsilon \left( \vec{r},\omega \right) =\varepsilon _{0}\left( x\right)
+\chi \left( \vec{r},\omega \right),
\label{eqn.7}
\end{equation}
where $\varepsilon _{0}\left( x\right) =1$ in vacuum, $\varepsilon
_{0}\left( x\right) =\varepsilon _{0}$ in the area of the grating
disposition, $\chi \left( \vec{r},\omega \right) $ is the space periodic
permittivity discribing the diffraction grating.

Let the permittivity $\chi \left( \vec{r},\omega \right) $ be the periodic
function of $y$ and $z$:
\[
\chi \left( \vec{r},\omega \right) =\sum_{\vec{\tau}\neq 0}\chi _{\tau
}\left( x\right) e^{-i\vec{\tau}\vec{\eta}},
\]
where $\vec{\eta}=y\vec{e}_{2}+z\vec{e}_{3}$ is the two-dimensional vector,
$\vec{e}_{2(3)}$ is the unit vector along $y(z)$ axis and
$\vec{\tau}=\tau _{y}\vec{e}_{2}+\tau _{z}\vec{e}_{3}$ is the
reciprocal lattice vector of the diffraction grating.

In the case $\chi =0$ the set of equations (1, 2)
desribes passing of the electromagnetic waves in the plain waveguide which
contains the layer of the matter. The dielectric susceptebility of the matter
is $\varepsilon _{0}$. The waveguide eigenmodes $\left| \vec{Y}_{n}\left(
x\right) \right\rangle $and eigenvalues $\kappa _{n}$ are well known \lbrack
18\rbrack . They can be used for simplifying the three-dimensional
Maxwell equations (1).

First of all, let us rewrite equation (1) as
\begin{eqnarray}
-\bigtriangleup \vec{E}\left( \vec{r},\omega \right) -\vec{\bigtriangledown}%
\left( \vec{\bigtriangledown}\left( \left( \varepsilon _{0}\left( x\right)
-1\right) \vec{E}\left( \vec{r},\omega \right) \right) \right) -\vec{%
\bigtriangledown}\left( \vec{\bigtriangledown}\left( \chi \left( \vec{r}%
\right) \vec{E}\left( \vec{r}\right) \right) \right) -
\label{eqn.8} \nonumber \\
-\frac{\omega ^{2}}{c^{2}}\left( \varepsilon _{0}\left( x\right) -1\right)
\vec{E}\left( \vec{r}\right) -\frac{\omega ^{2}}{c^{2}}\chi \left( \vec{r}%
\right) \vec{E}\left( \vec{r}\right) -\frac{\omega ^{2}}{c^{2}}\vec{E}\left(
\vec{r},\omega \right) = \\
=\frac{4\pi i\omega }{c^{2}}\left( \vec{j}(\vec{r},\omega )+\frac{c^{2}}{%
\omega ^{2}}\vec{\bigtriangledown}\left( \vec{\bigtriangledown}\vec{j}(\vec{r%
},\omega )\right) \right).\nonumber
\end{eqnarray}

Equations (8,6) allow us to find the electromagnetic
field $\vec E (\vec{r},\omega)$ radiated by electron beam. It
is well known that gain and the generation threshold can be find in the
linear approximation. In this case the beam current $\vec j$ is
the linear function of $\vec E (\vec r,\omega)$:
$\vec j= \vec j_0+\delta \vec j $, where $\vec{j}_{0}$ is the beam current
not perturbated by the radiated field,
$\delta \vec j \sim \vec E(\vec r,\omega) $
is the beam current induced by the radiated field.
In the linear approximation the set of movement equations (6) may be solved
by the following way: the electromagnetic field
$\vec E \left(\vec r_{\alpha }(t) ,\omega \right) $ in the right side
of equations (6) can be
represented as $\vec{E}\left( \vec{r}_{\alpha 0}+\vec{u}t,\omega \right) $,
where
$\vec{r}_{\alpha }\left( t\right) \simeq \vec{r}_{\alpha 0}+\vec{u}t$,
$\vec{r}_{\alpha 0}$ is the initial coordinate of the electron, $\vec{u}$
is the electron velocity;$\vec{v}%
_{\alpha }\left( t\right) \simeq \vec{u}$
in the absence of radiated field.

As a result, we can obtain from (6) that
\begin{eqnarray}
\delta \vec{v}_{\alpha }\left( \omega \right) =\frac{ie}{\omega m\gamma }%
\int \frac{d^{3}k^{\ \prime }}{\left( 2\pi \right) ^{3}}e^{i\vec{k}^{\
\prime }\,\vec{r}_{\alpha 0}}\left\{ \frac{\omega }{\omega +\vec{k}^{\
\prime }\vec{u}}\vec{E}\left( \vec{k}^{\ \prime },\omega +\vec{k}^{^{\prime
}}\vec{u}\right) \right. +
\label{eqn.9} \\
+\left. \left( \frac{\vec{k}^{\ \prime }}{\omega +\vec{k}^{\ \prime }\vec{u}}%
-\frac{\vec{u}}{c^{2}}\right) \left( \vec{u}\vec{E}\left( \vec{k}^{^{\prime
}},\omega +\vec{k}^{\ \prime }\vec{u}\right) \right) \right\}~,   \nonumber
\end{eqnarray}
\begin{equation}
\delta \vec{r}_{\alpha }\left( \omega \right) =\frac{i}{\omega }\delta \vec{v%
}_{\alpha }\left( \omega \right)~.
\label{eqn.10}
\end{equation}
The beam current induced by the radiated field is
\begin{eqnarray}
\delta \vec{j}\left( \vec{k},\omega \right) =\int e^{-i\vec{k}\vec{r}%
}e^{i\omega t}\delta j\left( \vec{r},t\right) d^{3}rd\omega =
\label{eqn.11} \\
=e\sum\limits_{\alpha }e^{-i\vec{k}\,\vec{r}_{\alpha 0}}
\left\{ \delta \vec{v}_{\alpha }( \omega -\vec{k}\vec{u})
-i\vec{u} \left( \vec{k}
\delta \vec{r}_{\alpha }( \omega -\vec{k}\vec{u}) \right)
\right\}.   \nonumber
\end{eqnarray}
After substitution of expressions (9, 10, 11)
in equation (\ref{eqn.8}) we shall
obtain the set of equations for the
field $\vec{E}\left( \vec{r},\omega \right) $.

Let us accomplish the Fourier
transformation of the field $\vec{E}\left( \vec{r},\omega \right) $:
\begin{equation}
\vec{E}\left( \vec{r},\omega \right) =\frac{1}{\left( 2\pi \right) ^{2}}\int
\vec{E}\left( x,\vec{k}_{\parallel }\right) e^{i\vec{k}_{\parallel
}\vec{\eta}}d^{2}k_{\parallel }.
\label{eqn.12}
\end{equation}
\vspace{1pt}
To obtain one-dimensional equation for the field
$\vec{E}(x,\vec{k}_{\parallel }) $ let us substitute expansion
(\ref{eqn.12}) in equation (\ref{eqn.8}).

Let $\chi=0$ (the smooth waveguide) and the electron beam is absent.
Then, equation (\ref{eqn.8})
allows us to find eigenfuctions
$\vec{Y}_{n}(x,\vec{k}_{\parallel})$ and eigenvalues
$\kappa_{n}^{2}(\vec{k}_{\parallel })$:
\begin{eqnarray}
-\frac{\partial ^{2}}{\partial x^{2}}\vec Y_n(x,\vec k_{\parallel})
-\vec{e}_{1}\frac{\partial }{\partial x}
\left[
\frac{\partial}{\partial x}
\left[
\left(\varepsilon _{0}(x)-1\right) Y_{nx}(x,\vec k_{\parallel})
\right]
+i\left(\varepsilon _{0}(x)-1\right)\vec{k}_{\parallel}
\vec Y_n(x,\vec k_{\parallel})
\right]
- \nonumber \\
-i\vec{k}_{\parallel}
\left[
\frac{\partial }{\partial x}
\left[
\left(\varepsilon _{0}(x) -1\right) Y_{x}(x,\vec k_{\parallel})
\right]
+i\left(\varepsilon _{0}(x)-1\right)\vec{k}_{\parallel}
\vec{Y}_{n}(x,\vec{k}_{\parallel})
\right] -
\label{eqn.13} \\
-\frac{\omega ^{2}}{c^{2}}\left( \varepsilon _{0}\left( x\right) -1\right)
\vec{Y}_{n}=\kappa _{n}^{2}\left( k_{\parallel }\right) \vec{Y}_{n}. \nonumber
\end{eqnarray}
If the vacuum-matter boundary  is sharp, expression (\ref{eqn.13})
gives the
well known equation for the waveguide containing the dielectric layer:

a) in vacuum
\begin{equation}
-\frac{\partial ^{2}}{\partial x^{2}}\vec{Y}_{n}=\kappa _{n}^{2}\vec{Y}_{n},
\label{eqn.14}
\end{equation}

b) in medium
\begin{equation}
-\frac{\partial ^{2}}{\partial x^{2}}\vec{Y}_{n}-\frac{\omega ^{2}}{c^{2}}%
\left( \varepsilon _{0}-1\right) \vec{Y}_{n}=\kappa _{n}^{2}\vec{Y}_{n}.
\label{eqn.15}
\end{equation}
Now we can decompose the field $\vec{E}(x,\vec{k}_{\parallel}) $
in terms of the waveguides eigenfunctions
$\vec{Y}_{n}(x,\vec{k}_{\parallel })$:
\begin{eqnarray}
\vec{E}\left( x,\vec{k}_{\parallel }\right) =\sum_{n}c_{n}\left( \vec{k}%
_{\parallel }\right) \left|
\vec{Y}_{n}(x,\vec{k}_{\parallel}) \right\rangle.
\label{eqn.16}
\end{eqnarray}
The field $\vec{E}(\vec{r},\omega)$ may be represented as
\begin{equation}
\vec{E}\left( \vec{r},\omega \right) =\frac{1}{(2\pi)^2}
\sum_{n}\int c_{n}(\vec k_{\parallel})
\left| \vec{Y}_{n}(x,\vec{k}_{\parallel}) \right\rangle
e^{i\vec{k}_{\parallel }\vec{\eta}}d^2 k_{\parallel}
\label{eqn.17}
\end{equation}
Let us substitute decomposition (16) into (8) and study the right side of (8)
(which is determined by the current $\vec{j}\left(
\vec{r},\omega \right) =\vec{j}_{0}\left( \vec{r},\omega \right) +\delta
\vec{j}\left( \vec{r},\omega \right)$) more attentively.
The set of equations (8) is a linear
system. As a result, we can omit the nonperturbative part of
current $\vec{j}%
_{0}$ and study (8) containing the induced current $\delta \vec{j}$ only.
Decomposition (16) allows us to obtain the
following expression for the right side of (8):
\begin{eqnarray}
M=\frac{4\pi i\omega }{c^{2}}
\int
\left\langle \vec{Y}_{n}(x,\vec{k}_{\parallel}) \right|
e^{-i\vec{k}_{\parallel}\vec{\eta}}
\left\{
\delta j\left( \vec{r},\omega \right) +\frac{c^{2}}{\omega ^{2}}%
\vec{\nabla}\left( \vec{\nabla}\delta \vec{j}(\vec{r},\omega)
\right)
\right\} dxd^2 \eta = \nonumber \\
=\frac{4\pi i\omega}{c^2}\frac{1}{(2\pi)^{3}}
\int\limits_{l}\int
\left\langle \vec{Y}_{n}(x,\vec{k}_{\parallel}) \right|
e^{i\vec{k}_{\parallel }\vec{\eta}}
\left\{
\delta j(\vec{k},\omega)-\frac{c^2}{\omega ^2}\vec{k}
\left( \vec{k}\delta \vec{j}(\vec k,\omega) \right)
\right\}
e^{i\vec k \vec r}d^{3}kdx=
\label{eqn.18} \\
=\frac{4\pi i\omega}{c^2}\frac{1}{2\pi }
\int\limits_{l}\int
\left\langle \vec{Y}_{n}(x,\vec{k}_{\parallel }) \right|
\left\{
\delta \vec j (\vec k,\omega) -\frac{c^2}{\omega^2}\vec{k}
\left(\vec{k}\delta \vec j (\vec{k},\omega) \right)
\right\} e^{ik_x x}dxdk_x ,  \nonumber
\end{eqnarray}
where $\vec{k}=\left( k_{x},\vec{k}_{\parallel }\right) $.
It should be mentioned that the electron beam current density
$\vec j (\vec r,\omega)$ is not equal to zero only in the vacuum area
restricted by the
beam transverse size $l$. This fact results in the appearance of
integral $\int\limits_{l}$ in (18) which means the integration over the
area where
$\vec j (\vec r,\omega)\neq 0$ (see Fig.3,4).
As a result, we have
\begin{equation}
M=\frac{4\pi i\omega }{c^2}\frac{1}{2\pi }
\int dk_{x}
\left\langle \vec Y_n(k_x,\vec{k}_{\parallel}) \right|
\left\{
\delta \vec j (\vec k,\omega) -\frac{c^2}{\omega ^2}\vec{k}
\left(\vec k \vec \delta j(\vec k,\omega) \right)
\right\},
\label{eqn.19}
\end{equation}
where
\[
\left\langle \vec{Y}_n(k_x,\vec k_{\parallel})\right|
=\int\limits_{l}
\left\langle \vec{Y}_n(k_x,\vec{k}_{\parallel}) \right|
e^{ik_x x}dx
\]
The expression for the current density
$\delta \vec j (\vec{k},\omega) $
contains the sum
$F=\sum\limits_{\alpha }e^{-i(\vec{k}-\vec{k}^{\prime}) \vec{r}_{\alpha 0}}$.
Let us average this sum over distribution of the particles in the beam:
\begin{equation}
\sum\limits_{\alpha }e^{i\left( \vec{k}-\vec{k}^{\prime }\right) \vec{r}%
_{\alpha \,0}}\simeq \Phi \left( k_{x}-k_{x}^{\prime }\right) \left( 2\pi
\right) ^{3}n_{0}\delta \left( \vec{k}_{\parallel }-\vec{k}%
_{\parallel }^{\prime }\right),
\label{eqn.20}
\end{equation}
where $\Phi(k_{x}-k_{x}^{\prime }) =\frac{1}{2\pi }
\int\limits_{a}^{b}e^{-i(k_{x}-k_{x}^{\prime }) x}\varphi(x) dx$,
$\frac 1l \int\limits_{a}^{b} \varphi(x)dx=1$,
$l$ is the characteristic transversal size of
the beam, function $\varphi(x)$ describes the distribution of the
particles along $x$-direction, $n_{0}$ is the electron density of the beam.

Using (20) we can write $\delta j (\vec k,\omega)$ as
\begin{eqnarray}
\delta j (\vec{k},\omega) =
\vec{u}\delta \varphi(\vec k,\omega)
=\vec{u}\frac{ie^2 n_0}{\omega c^2
(\omega -\vec{k}\vec{u})^2 m\gamma}
\frac{1}{2\pi}
\int\limits_{l}dx^{\prime}
e^{-ik_{x}x^{\prime }}\varphi(x^{\prime }) \times
\label{eqn.21} \\
\times \left( -ik_{x}\frac{\partial }{\partial x^{\prime }}c^{2}+\frac{%
c^{2}-u^{2}}{u^{2}}\omega ^{2}\right)
\left(
\vec u \vec E(x^{\prime },\vec{k}_{\parallel},\omega)
\right).  \nonumber
\end{eqnarray}
As a result, using (21) we can represent (19) as
\begin{eqnarray}
M=
\frac{4\pi i\omega}{c^2} \frac 1{2\pi}
\int dk_x
\left\langle \vec Y_n(k_x,\vec k_{\parallel})\right|
\left\{
\vec u - \frac {c^2}{\omega^2}\vec k(\vec k,\vec u)
\right\}
\delta \varphi(\vec k,\omega)=
\label{eqn.22} \\
=\frac{4\pi i\omega }{c^2} \frac{1}{2\pi }
\int dk_x
\left\langle \vec E_{n}\left( k_{x},\vec{k}_{\parallel }\right) \right|
\left\{
\vec{u}-\frac{c^2}{\omega ^2}\vec{k}
\right\}
\delta \varphi(\vec{k},\omega)  \nonumber
\end{eqnarray}
In the cold beam case
(when the condition $k_{x}u_{x}\frac{1}{c}\ll L$ to be fulfilled)
we can write
\begin{eqnarray}
M=
\frac{4\pi i\omega }{c^2}
\frac{ie^2 n_0}{(\omega -\vec{k}_{\parallel}\vec{u})^2 m\gamma \omega c^2}
\frac{1}{2\pi }
\int\limits_{l}dx
\left\langle \vec{Y}_n(x,\vec k_{\parallel},\omega) \right|
\left(
\vec u-\frac{c^2}{\omega}\vec{k}_{\parallel}
-i{\hat {\frac \partial {\partial x}}}\frac{c^2}{\omega }\vec{e}_1
\right) \times
\label{eqn.23} \\
\times \left(
\hat {\frac{\partial }{\partial x}}\phi(x){\frac{\partial}{\partial x}}c^2+
\frac{c^2 \omega ^2}{u^2}\frac{1}{\gamma ^2}\right)
\left( \vec u \vec E (x^{\prime},\vec{k}_{\parallel},\omega)
\right),  \nonumber
\end{eqnarray}
where operator $\hat {\frac{\partial }{\partial x}}$ acts on the functions
disposed on its left.

Substituting the decomposition
$\vec{E}(x^{\prime},\vec{k}_{\parallel},\omega) =
\sum\limits_n^{\prime}
c_{n^{\prime}}(\vec{k}_{\parallel})
\left| \vec{Y}_{n}(x,\vec{k}_{\parallel}) \right\rangle $
into (23) and using the ortogonality of the eigenfunctions we obtain
from (8, 23)
\begin{eqnarray}
\left( k_{\parallel }^{2}-(\frac{\omega ^{2}}{c^{2}}-\kappa
_{n}^{2}) \right) c_{n}(\vec{k}_{\parallel}) -
\frac{\omega^2}{c^2}
\sum\limits_{\vec \tau,n^{\prime }}
\chi_{eff}^{nn^{\prime }}(\vec{k}_{\parallel},\vec{k}_{\parallel}+\vec{\tau})
c_{n^{\prime }}(\vec{k}_{\parallel}+\vec{\tau})
=
\label{eqn.24} \\
=\frac{4\pi i\omega }{c^2}
\sum\limits_{n^{\ \prime }}\frac{%
ie^{2}n_{0}A_{nn^{\prime }}}{\left( \omega -\vec{k}_{\parallel}\vec{u}
\right)^2 m\gamma \omega} c_{n^{\prime }}( \vec{k}_{\parallel}),  \nonumber
\end{eqnarray}
where
\begin{eqnarray}
A_{nn^{\prime }}=\frac{1}{2\pi c^2}
\int\limits_{l}dx
\left[
\left\langle \vec{Y}_{n}( x,\vec{k}_{\parallel },\omega)\right|
\left\{ \vec{u}-\frac{c^2}{\omega }\vec{k}_{\parallel}
-i(\hat {\frac{\partial }{\partial x}}\frac{c^2}{\omega }\vec{e}_1)
\right\} \right] \times \nonumber\\
\times \left[
\hat {\frac{\partial }{\partial x}}\varphi(x)
{\frac{\partial }{\partial x}}c^2+
\frac{c^2 \omega ^2}{u^2 \gamma ^2}\right]
\left(
\vec u \left| \vec Y_n^{^{\prime }}(x,\vec k_{\parallel},\omega)\right\rangle
\right)
\approx
\frac 1{2\pi c^2}
\int\limits_a^b dx
\left[
\left\langle \vec{Y}_{n}(x)\right|
\left( \vec{u}-\frac{c^2}{\omega}\vec{k}_{\parallel}\right)
\right]
\times \\
\label{eqn.25}
\times
\left( \frac{\partial\varphi}{\partial x}\frac{\partial}{\partial x}c^2\right)
\left( \vec{u}\left| \vec{Y}_n(x) \right\rangle
\right), \nonumber
\end{eqnarray}
where
$\chi _{eff}^{nn^{\prime }}(\vec{k}_{\parallel },\vec{k}_{\parallel}
+\tau)$ is the effective permittivity. It contains
two terms: the first proportional to
$\chi( \vec{r},\omega)$ and the second
proportional to
$\vec{\nabla}\left( \vec{\nabla}\chi( \vec{r},\omega) \right)$:
\begin{eqnarray}
\chi _{eff}^{nn^{\prime }}
(\vec{k}_{\parallel},\vec{k}_{\parallel}+\vec \tau)
=\int dx
\left\langle \vec{Y}_n( x,\vec{k}_{\parallel})\right|
\chi_{\tau }(x)
\left|
\vec{Y}_{n^{\prime }}(x,\vec{k}_{\parallel }+\vec \tau) \right\rangle +
\label{eqn.26} \\
\int dx
\left\langle \vec{Y}_{n}(x,\vec{k}_{\parallel })\right|
\frac{c^2}{\omega ^2}
\hat{\vec{k}}\left( \hat{\vec{k}}\chi_{\tau }(x)
\left| \vec{Y}_{n^{\prime }}(x,\vec{k}_{\parallel }+\vec \tau) \right\rangle
\right),  \nonumber
\end{eqnarray}
where
$\hat{\vec{k}}=\frac{\partial }{\partial x}\vec e_1+i\vec k_{\parallel}$.

Let the condition $\vec{k}_{\parallel }\gg \frac{2\pi }{d}$ be
fullfiled ($d$ is the diametrical size of the waveguide). In this case the
second term of (26) is less than the first one.

We can simplify the system (2) solving by assuming a practically important
case of
a single mode $n$ existance in the waveguide.
It is possible when condition
$\omega ^2/c^2 \chi^{nn^{\prime}}
(\vec k_{\parallel},\vec k_{\parallel}+\vec{\tau})
\left( \kappa _{n}^{2}-\kappa_{n^{\prime }}^{2}\right) ^{-1}\ll 1$
is fulfilled. In this case all terms with $n^{\prime }\neq n$ in the sum on
$n^{\prime }$ in (24) can be omitted.

As a result, we obtain the set of equations which is similar to that
describing the multiwave dynamical diffraction of
electromagnetic waves in a diffraction grating.

In the two wave diffraction case the Bragg condition
accomplishes for two waves with wavevectors $\vec{k}_{\parallel }$
and $\vec{k}_{\parallel }^{\prime }$: $\vec{k}_{\parallel}^{\prime }\simeq
\left( \vec{k}_{\parallel }+\vec{\tau}\right) $,
$\left| \vec k_{\parallel}^{\prime}\right| \simeq
\left| \vec k_{\parallel}\right| $ and the set of equations
(\ref{eqn.24}) can be written as
\begin{eqnarray}
\left[
k_{\parallel }^2-\frac{\omega ^2}{c^2}\varepsilon _{0}+%
\frac{\omega _{L}A_{nn}}{\gamma c^{2}\left( \omega -\vec{k}_{\parallel }%
\vec{u}\right) ^{2}}
\right]
c_{n}( \vec{k}_{\parallel}) -
\frac{\omega ^2}{c^2}\chi_{eff}^{nn}( \vec{k}_{\parallel },\vec{k}%
_{\parallel}+\vec{\tau})
c_{n}( \vec{k}_{\parallel }+\vec{\tau}) =0,
\label{eqn.27} \\
-\frac{\omega ^2}{c^2}
\chi _{eff}^{nn}( \vec{k}_{\parallel }+\vec{\tau},\vec{k}_{\parallel })
c_{n}(\vec{k}_{\parallel }) +
\left[
(\vec{k}_{\parallel}+\vec{\tau})^2-\frac{\omega ^2}{c^2}\varepsilon_0
\right]
c_{n}(\vec{k}_{\parallel }+\vec{\tau}) =0,  \nonumber
\end{eqnarray}
where $\varepsilon _0=1-c^2 \kappa _n^2 / \omega^2$, $\omega _{L }$ is the
Lengmuer frequency of the electron beam ($\omega_L^2=4 \pi e^2 n_0 / m$).

The set of equations (27) is similar to that for the
electromagnetic field amplitudes describing lasing in volume
FEL for the case of beam moving in volume diffraction grating \lbrack 6\rbrack.
The main discrepancy appeares in the dependence of the equations (27) on
$\vec{k}_{\parallel}$. The similar set of equations for volume FEL
depends on $\vec{k}$.

As a result, we can conclude that all the  main results obtained for
VFELs  holds true for the vacuum VFEL. First of all,
the non-trivial solution of system (27) exists only
when the determinant of the system is
equal to zero. This allows us to obtain the dispersion equation for
$\vec{k}_{\parallel }$ and $\omega$:
\begin{eqnarray}
\left( \omega -\vec{k}_{\parallel }\vec{u}\right) ^{2}\left[ \left(
k_{\parallel }^{2}c^{2}-\omega ^{2}\varepsilon _{0}\right) \left(
\left( \vec{k}_{\parallel }+\vec{\tau}\right) c^{2}-\omega
^{2}\varepsilon _{0}\right) -\omega ^{4}\chi _{\tau }^{nn}\chi _{-\tau
}^{nn}\right] =
\label{eqn.28} \\
=-\frac{\omega _{L}^{2}}{\gamma }A_{nn}\left( \left( \vec{k}_{\parallel
}+\vec{\tau}\right) ^{2}c^{2}-\omega ^{2}\varepsilon _{0}\right)  \nonumber
\end{eqnarray}
\vspace{1pt}

According to \lbrack 5-8\rbrack\, the study of the dispersion equation
let us find the condition of the appearance of the convection and absolute
instability of the beam and, as a result, obtain the gain and the generation
threshold.

In the plain waveguide case ($\chi _{\tau}^{nn}=0$) from (28) we have
\begin{equation}
\left( \omega -\vec{k}_{\parallel }\vec{u}\right) ^{2}\left(
k_{\parallel }^{2}c^{2}-\omega ^{2}\varepsilon _{0}\right) =-\frac{%
\omega _{L}^{2}}{\gamma }A_{nn}
\label{eqn.29}
\end{equation}
\begin{equation}
\left( \omega -\vec{k}_{\parallel }\vec{u}\right) ^{2}\left(
k_{\parallel }c-\omega \sqrt{\varepsilon _{0}}\right) \simeq -\frac{%
\omega _{L}^{2}}{2\omega \sqrt{\varepsilon _{0}}\gamma }A_{nn}
\label{eqn.30}
\end{equation}

Equations (\ref{eqn.29}), (\ref{eqn.30}) coincide with the
equations describing the wave
spectrum for Cherenkov instability of the beam in medium. From
(\ref{eqn.30}) we can
obtain that $k_{\parallel }$ has imaginary part $Imk_{\parallel}$
and
\begin{equation}
Im k_{\parallel}=
\frac {\sqrt{\varepsilon_0}}{2c}
\left( {\frac{\omega _{L}^{2}\left|A_{nn}\right|}
{ 2 \omega \varepsilon_0 \gamma }}  \right)^{\frac13},
\end{equation}
when the Cherenkov condition $1-\omega / c \sqrt{\varepsilon_0} cos \vartheta=0$
is fulfilled.
As we see, $Imk_{\parallel }$ is proportional to $n_{0}^{1/3}$ ($n_{0}$ is the
density of the electron beam). It means that the gain is proportional to
$n_{0}^{1/3}$ as well. This dependence is typical for all types of
one-dimensional FEL in the collective regime \lbrack 3\rbrack .

Let the waveguide contains a diffraction grating
$\left( \chi _{\tau}^{nn}\neq 0\right) $.
The wave spectrum is described by equation (\ref{eqn.28}). When
the coefficient $A_{nn}=0$, equation (\ref{eqn.28}) splites into two
equations:
\begin{equation}
\left( k_{\parallel }^{2}c^{2}-\omega ^{2}\varepsilon _{0}\right)
\left( \left( \vec{k}_{\parallel }+\vec{\tau}\right) ^{2}c^{2}-\omega
^{2}\varepsilon _{0}\right) -\omega ^{4}\chi _{\tau }^{nn}\chi _{-\tau
}^{nn}=D\left( \vec{k}_{\parallel },\omega \right) =0
\label{eqn.31}
\end{equation}
\begin{equation}
\left( \omega -\vec{k}\vec{u}\right) ^{2}=0
\label{eqn.32}
\end{equation}
Equation (\ref{eqn.31}) describes the electromagnetic wave spectrum for the
waiveguide containing a diffraction grating. Equation (\ref{eqn.32})
describes the
wave spectrum of the electron beam charge density. Let us study the
solutions of (\ref{eqn.28}) near the point where the left side of
(\ref{eqn.28}) is equal to zero.
The solution
of (31, 32) in the vicinity of the exact Bragg condition
$\left| \vec{k}_{\parallel }+\vec{\tau}\right| \simeq
k_{\parallel }$ can be written in the form
\begin{equation}
k_{z0}=k_{z}^{0}\left( 1+\delta \right) ,\quad \omega _{0}=k_{z}^{0}u\left(
1+\delta \right),
\label{eqn.33}
\end{equation}
where $\delta \ll 1$ and $k_{z}^{0}$\ can be found from exact Bragg
conditions
\begin{equation}
k_{z}^{0}=-\frac{2k_{y}\tau _{y}+\tau ^{2}}{2\tau _{z}},
\label{eqn.34}
\end{equation}
$z$ axis is parallel to the beam velocity $\vec{u}$.
From (28,34) we can obtain for $\delta $
\begin{eqnarray}
\delta =\frac{\chi _{\tau }^{nn}\chi _{\tau }^{nn}-\left( \eta +\xi \right)
^{2}}{2\nu \left( \eta +\xi \right) };
\label{eqn.35} \\
\nu =\frac{\tau _{z}}{k_{0z}},\eta =\frac{k_{y}^{2}}{k_{z}^{02}},\xi
=1-\beta ^{2}\varepsilon _{0}  \nonumber
\end{eqnarray}
Now we can study equations (\ref{eqn.27}). Let us rewrite (\ref{eqn.27}) as
\begin{equation}
\left( \omega -k_{z}u\right) ^{2}D\left( k_{z},\omega \right) =A\left(
k_{z},\omega \right),
\label{eqn.36}
\end{equation}
where
\begin{equation}
A\left( k_{z},\omega \right) =-\frac{\omega _{L}^{2}}{\gamma }A_{nn}\left(
\left( \vec{k}_{\parallel }+\vec{\tau}\right) ^{2}c^{2}-\omega
^{2}\varepsilon _{0}\right)
\label{eqn.37}
\end{equation}
The solution of (\ref{eqn.36}) can be represented as
\begin{equation}
\omega =\omega _{0}+\omega ^{\prime },\quad k_{z}=k_{zo}+k_{z}^{\prime }
\label{eqn.38}
\end{equation}
where $\left| \omega ^{\prime }\right| \ll \omega _{0}$ and $\left|
k_{z}^{\prime }\right| \ll \left| k_{z0}\right| $.

Let us write $D\left( k_{z},\omega \right) $ in the form
\begin{eqnarray}
D\left( k_{z},\omega \right) =\left( \frac{\partial D}{\partial \omega }%
\right) _{\omega _{0},k_{z0}}\omega ^{\prime }+\left( \frac{\partial D}{%
\partial k_{z}}\right) _{_{\omega _{0},k_{z0}}}k_{z}^{\prime }+\frac{1}{2}%
\left( \frac{\partial ^{2}D}{\partial k_{z}^{2}}\right) _{_{\omega
_{0},k_{z0}}}k_{z}^{\prime ^{2}}+
\label{eqn.39} \\
+\frac{1}{2}\left( \frac{\partial ^{2}D}{\partial \omega ^{2}}\right)
_{\omega _{0},k_{z0}}\omega ^{\prime ^{2}}+\left( \frac{\partial ^{2}D}{%
\partial \omega \partial k_{z}}\right) _{\omega _{0},k_{z0}}\omega ^{\prime
}k_{z}^{\prime }+...  \nonumber
\end{eqnarray}
It can be shown that $\left( \partial D / \partial \omega \right)
_{\omega _{0},k_{z0}}$ can not be equal to zero.
That is why, we can omit the term
proportional to $\left( \partial ^{2}D / {\partial \omega ^{2}}\right)
_{\omega _{0},k_{z0}}$. On the other hand, the derivative
$\left({\partial D} / {\partial k_{z}}\right) _{_{\omega _{0},k_{z0}}}$
can be equal to zero. In this case equation (36) can be written as
\begin{equation}
\left( \omega ^{\prime }-k_{z}^{\prime }u\right) ^{2}\left( k_{z}^{\
\prime ^{2}}-F\omega ^{\prime }\right) =\frac{2A\left( k_{zo},\omega
_{0}\right) }{\left(
\frac{\partial ^{2}D}{\partial k_{z}^2}\right) _{_{\omega_{0},k_{z0}}}},
\label{eqn.40}
\end{equation}
where
\[
F=-2\left( \frac{\partial D}{\partial \omega }\right) _{_{_{\omega
_{0},k_{z0}}}}\left( \frac{\partial ^{2}D}{\partial k_{z}^{2}}\right)
_{_{\omega _{0},k_{z0}}}^{-1}
\]
For $\omega ^{\prime }\rightarrow 0$ equation (40) takes form
\begin{equation}
k_{z}^{\prime ^{4}}=\frac{2A}{u}\left( \frac{\partial ^{2}D}{\partial
k_{z}^{2}}\right) _{_{\omega _{0},k_{z0}}}^{-1}
\label{eqn.41}
\end{equation}
As a result, equation (\ref{eqn.41}) gives
\begin{equation}
Imk_{z}=\left( \frac{\omega _{0}\left| Q\right| \left| \chi _{\tau }\right|
\tau }{u^{2}c^{2}\sqrt{2\tau _{z}^{2}-\tau ^{2}}}\right) ^{1/4}=\left( \frac{%
\omega _{L}^{2}A_{nn}\left| \chi _{\tau }\right| \tau }{2\gamma u^{2}c^{2}%
\sqrt{2\tau _{z}^{2}-\tau ^{2}}}\right) ^{1/4}
\label{eqn.42}
\end{equation}

Let us remind that the Langmuer frequency $\omega _{L}\sim n_{0}^{1/2}$. So,
accordingly to (\ref{eqn.42}),
we have been obtained a very important result:
in vacuum VFEL both $Imk_{z}$ and
the gain are proportional to $n_{0}^{1/4}$ for the definite orientation of the
diffraction grating in the waveguide  in contrast with the conventional
one-dimensional FEL for which the gain is proportional to $n_{0}^{1/3}$.
From (32) and (43) we have:
\begin{equation}
\frac {Imk_z}{Im k_{\parallel} } \approx
\left( \frac{\omega _L^2 \left| A_{nn} \right| }
{ \omega^4 \chi_{\tau}^3 \gamma }\right) ^{-\frac {1}{12}} \gg 1,
\label{eqn.44}
\end{equation}
because
$\frac{\omega _L^2 \left| A_{nn} \right| }{ \omega^4 \chi_{\tau}^3 \gamma }
\ll 1$,
$\frac{\omega_L^2}{ \omega^2}\ll 1$ and $ A_{nn} \ll \omega^2$ .
As a result, in our case of the volume feedback
the gain is larger then that for one-dimensional feedback.
For example, the dependence  of the threshold current density for the
volume and one-dimensional geometries on the length is represented  in Fig.7.

\begin{figure}[h]
\epsfxsize = 7.8 cm
\centerline{\epsfbox{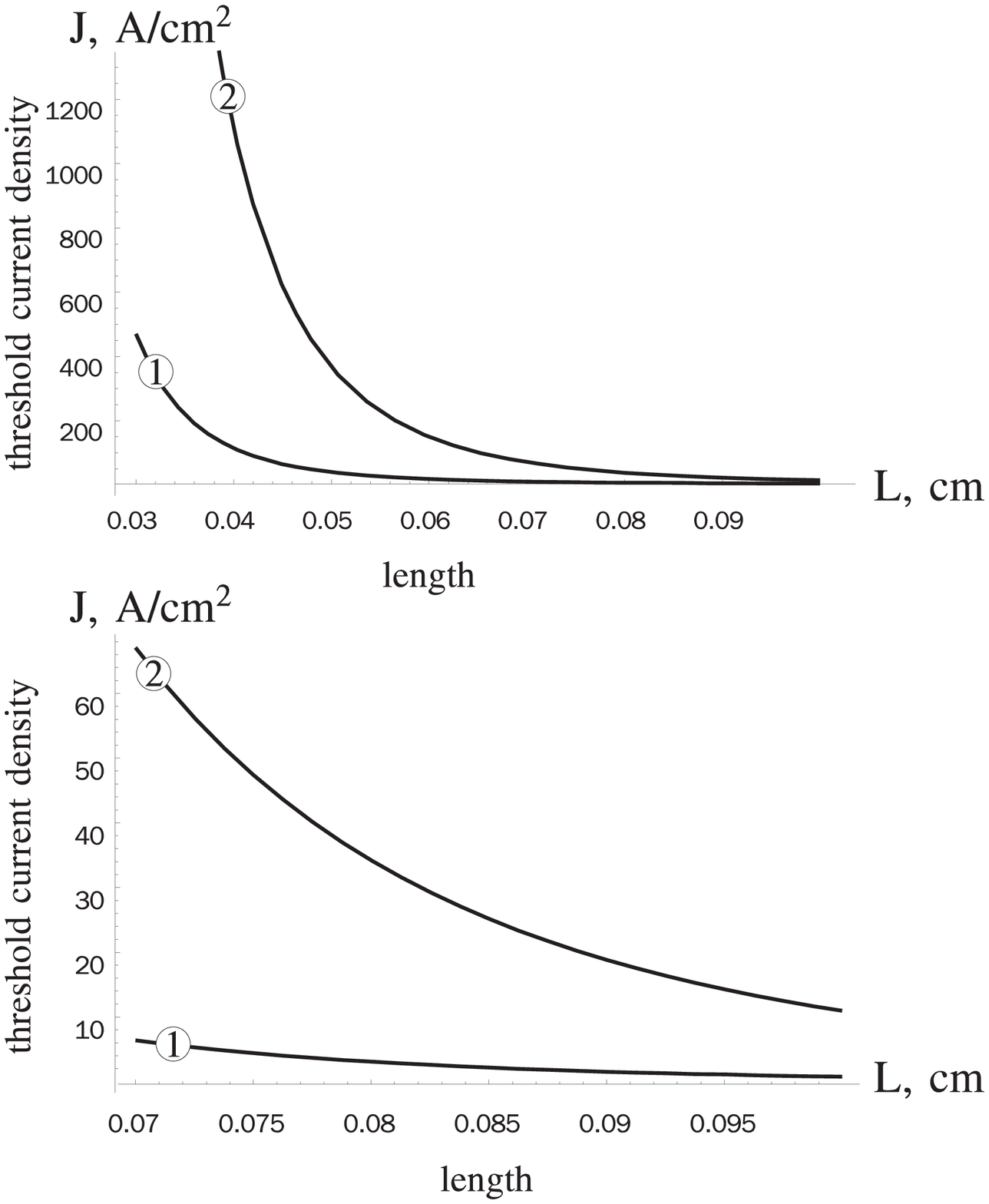}}
\caption{The dependence  of the threshold current density for the
volume (1) and one-dimensional (2) geometries on the length
($\lambda=6283 \stackrel{\circ}{A}$).}
\end{figure}

The gain becomes higher when the distributed
feedback is formed by the multi-wave dynamical diffraction:
\begin{equation}
Im k_z \approx
\frac{\omega}{c} \chi_\tau
\left( \frac{ \omega _L^2 \left| A_{nn} \right| }
{\omega^4 \chi_{\tau}^3 \gamma }\right)^{\frac {1}{S+1}},
\label{eqn.43}
\end{equation}
where $S$ is the number of diffracted waves.

\begin{center}
{\bf CONCLUSION}
\end{center}

The vacuum VFEL amplification and  the generation process develop
more intensively
then in ordinary FEL using one-dimensional distributed feedback.
Such the VFEL, if
realised, could be made with much more compact device structure compared with
the FEL and therefore, may be interesting for applications in different
wavelength ranges:
from submillimeter to $X$-ray. Such the VFEL can be realised
on the basis of the existing accelerators.

Author is gratefull to K. Batrakov for the help in the threshold
current densities
computer simulation carrying out.

\begin{center}
{\bf REFERENCES}
\end{center}

\lbrack 1\rbrack\ Jh.C.Marshall Free-Electron Lasers (London: Macmillan,
1984).

\lbrack 2\rbrack\ Free-Electron Lasers 1996. Proccedings of the Eighteenth
International Free Electron Lasers Conference Rome Italy, August 26-31,
1996, North-Holland 1997.

\lbrack 3\rbrack\ A.Gover, Z. Livni  Optics Communications
{\bf26} (1978), 375.

\lbrack 4\rbrack\ J.R.Pierci. Travelling wave tubes (Van Nostrand, Princeton
1950).

\lbrack 5\rbrack\ V.G.Baryshevsky and I.D.Feranchuk. Phys. Lett.
{\bf A102} (1984), 141.

\lbrack 6\rbrack\ V.G.Baryshevsky, K.G.Batrakov, I.Ya.Dubovskaya. Journ.
Phys. D: Appl. Phys. {\bf 24} (1991), 1250.

\lbrack 7\rbrack\ V.G.Baryshevsky, K.G.Batrakov, I.Ya.Dubovskaya. NIM
{\bf A358} (1995), 493.

\lbrack 8\rbrack\ V.G.Baryshevsky, K.G.Batrakov, I.Ya.Dubovskaya. NIM
{\bf A358} (1995), 508.

\lbrack 9\rbrack\ V.G.Baryshevsky, K.G.Batrakov, I.Ya.Dubovskaya  Free
Electron Lasers 1996, Elsevier Science, (1997).

\lbrack 10\rbrack\ V.G.Baryshevsky  Docklady Academy of Science of the USSR
{\bf 299} N6, (1988), 1363.

\lbrack 11\rbrack\ S.J.Smith and E.M.Purcell. Phys. Rev.
{\bf 92} (1953), 1069.

\vspace{1pt}\lbrack 12\rbrack\ K.J.Woods, J.E.Walsh, R.E.Stoner and at.al.
Phys. Rev. Lett. {\bf 74} (1995), 3809.

\lbrack 13\rbrack\ F.S.Rusin and G.D.Bogomolov JETP Lett.
{\bf 4} (1966), 160.

\lbrack 14\rbrack\ Richard P.Leavit, Donald E.Wortman and Clyde A.Morrison.
Appl. Phys. Lett.
{\bf 35} (1979), 363.

\lbrack 15\rbrack\ V.P.Shestopalov. Diffraction Electronics, VS. Joint
Publications Service, April 1978.

\lbrack 16\rbrack\ J.M.Wachel J.Appl. Phys.
{\bf 50} (1979), 49.

\lbrack 17\rbrack\ J.Urata, M.Goldstein, M.F. Kimmitt, A. Naumov, C.Platt
and J.E.Walsh Phys.Rev.Lett
{\bf 80} (1998), 516.

\lbrack 18\rbrack\ A.Yariv. Quantum Electronics, 2-nd ed. (Wiley, New York,
1975).
\end{document}